\documentstyle[graphicx,aps,amstex]{revtex}
\newcommand{\vu}{{\mathbf{v}}}

\newcommand{\er}{{\mathbf{r}}}
\begin{document}
\twocolumn
\draft
\title{Collective oscillations of an interacting trapped Fermi gas}
\author{L.\ Vichi and S.\ Stringari} \address{Dipartimento di Fisica,
Universit\`a di Trento and Istituto Nazionale per la Fisica della
Materia, I-38050 Povo,  Italy}
\date{May 12, 1999} \maketitle
\begin{abstract}

We calculate the effects of two-body interactions on the low frequency 
oscillations 
of a  normal  Fermi gas confined in a harmonic trap.
The mean field contribution to the collective frequencies is evaluated
in the collisionless regime 
using a sum rule approach. We also discuss 
the transition between the collisionless and hydrodynamic regime
with special emphasis to the spin dipole mode in which 
two atomic clouds  occupying different spin states oscillate
in opposite phase. The spin dipole mode is predicted to be overdamped in 
the hydrodynamic regime. The relaxation time is calculated
as a function of temperature and the effects of Fermi statistics
are explicitly pointed out. 

\end{abstract}


The investigation of collective excitations in trapped atomic gases
has become an active research field, stimulated by the experimental
realization of Bose-Einstein condensation \cite{bec}. 
Because of the
high density of the condensate, interaction effects in 
cold Bose gases are
crucial \cite{rmp} in order to explain the experimental results
 for the collective frequencies \cite{exp}.
In Fermi gases the Pauli exclusion principle makes the density of
the trapped gas more dilute, thereby reducing the effects of
interactions. The equilibrium  properties
of trapped Fermi gases as well as of Fermi-Bose mixtures have been
already the object of theoretical calculations based on mean 
field approaches \cite{houbiers,theo-fermi,molmer,tnpi1}. On the other hand
experiments aiming to produce samples of Fermi gases in
conditions of quantum degeneracy are also becoming available \cite{exp-fermi}.

In this letter we provide a theoretical discussion of
interaction effects on the collective oscillations of a trapped Fermi
gas. The very high precision of frequency
measurements, the large value of the scattering length exhibited by
some atomic species and the possibility of pointing out effects of
quantum statistics in the collisional term, 
make the study of collective oscillations in these
systems a promising area of investigation.
We will consider Fermi gases
occupying two distinct spin states (hereafter called, for simplicity,
spin up and spin down respectively). In fact only in this case 
can the effects of the interaction generated by $s$-wave
scattering be explored. We will further limit the discussion to the case of
excitations of low multipolarity in the normal  phase, corresponding to the 
easiest realization of future
experiments. In fact the transition to the superfluid phase is
predicted \cite{kagan,stoof} to occur at very low temperatures. For a
discussion of the oscillations in the superfluid phase
see \cite{stoof,petrov}. 

At very low temperatures collisions are quenched by
Fermi statistics and the system is in the collisionless
regime, corresponding to the propagation of zero sound in
traditional Fermi liquids \cite{pines}.
A useful approach to the study of collective excitations in this
regime is provided by  sum rules \cite{lipparini}.
In this approach the excitation energies are estimated through the ratio
\begin{equation}
\hbar \omega = \sqrt{\frac{m_3}{m_1}}
\end{equation}
between the  $k=3$ and $k=1$  moments
$m_k=\int S(E) E^kdE$ of the dynamic structure
factor
\begin{equation}
S(E)=\sum_n|\langle n|F|0 \rangle|^2\delta(E-\hbar \omega_n)
\end{equation}
relative to a given operator $F$. For simplicity we
have considered the $T=0$ case, 
but the formalism can be naturally extended to
finite temperature. The moments $m_1$ and $m_3$ are easily
evaluated in terms of commutators. In fact, using the completeness
relation, one can write
\begin{align} \label{m1}
m_1=  & \frac 1 2 \langle
0|[F^{\dagger},[H,F]]|0 \rangle,   \\ \label{m3}
m_3= & \frac 1 2 \langle 0
|[[F^{\dagger},H],[H,[H,F]]]|0 \rangle ,
\end{align}
thereby avoiding the most
difficult problem of determining the full function $S(E)$. In
eq.(\ref{m1}-\ref{m3}) $|0 \rangle$ is the ground state of the many
body system. 
Let us consider a symmetric configuration with 
$N_{\uparrow}=N_{\downarrow}= N/2$
and the same confining potential
$V_{ext}^{\uparrow}=V_{ext}^{\downarrow}=V_{ext}$ for the two spin species.
We will evaluate the commutators using the mean field hamiltonian
\begin{equation} \label{H}
H = \sum_i \frac{p_i^2}{2m}+V_{ext}+g\sum_{i\uparrow,j\downarrow}
\delta( {\bf{r}} _i -{\bf{r}} _j)
\end{equation}
where $g=4 \pi \hbar^2 a/m$ is the interaction coupling constant fixed
by the $s$-wave scattering length $a$. By choosing 
an isotropic harmonic potential 
$V_{ext}(r)=\frac 1 2 m \omega_{ho}^2 r^2$
the commutators with the monopole ($F=\sum_ir_i^2$),
quadrupole ($F=\sum_i r_i^2-3z_i^2$) and dipole ($F=\sum_i z_i$)
operators can be easily evaluated and   one obtains the
following  results \cite{ss} 
\begin{equation} \label{M}
 \omega_M = 
\sqrt{\frac{2E_{kin}+6E_{ho}+6E_{int}}{Nm\langle r^2 \rangle}}= 
2\omega_{ho} \sqrt{1+\frac 3  8 \frac{E_{int}}{E_{ho}}}
\end{equation}
\begin{equation} \label{Q}
\omega_Q =   \sqrt{\frac{E_{kin}+E_{ho}}{Nm\langle r^2 \rangle}}=2
 \omega_{ho} \sqrt{1-\frac 3 4 \frac{E_{int}}{E_{ho}}}
\end{equation}
\begin{equation} \label{D}
 \omega_D =  \omega_{ho}
\end{equation}
where $E_{int}=g\int \rho(r)^2d\er /4$ is the mean field interaction energy,
$E_{ho}=m\omega_{ho}^2\int r^2\rho(r)d\er /2$ is the potential
oscillator energy,  
$E_{kin}=\sum_i\langle 0| p_i^2/2m |0 \rangle$ is the kinetic energy
relative to the ground state and $\rho = \rho_{\uparrow} + \rho_{\downarrow}$ 
is the total density of the gas.
In deriving the second equalities in (\ref{M}-\ref{Q}) we have used the
virial theorem \cite{rmp}
\begin{equation} \label{virial}
2E_{kin}-2E_{ho}+3E_{int} =0
\end{equation}
which allows one to calculate the deviations of the collective frequencies
from the ideal gas prediction $2\omega_{ho}$ in terms of the ratio
  $E_{int}/E_{ho}$. 
Differently from the quadrupole and the monopole, the dipole frequency is not
affected by interactions. In fact this mode 
corresponds to the oscillation of the center of mass of the gas driven
by the external harmonic potential. 
It is worth noticing that results (\ref{M}-\ref{D}) hold also in the
case of trapped bosons \cite{ss}, provided one uses  the 
corresponding expression for the interaction energy. 
In the case of Bose-Einstein
condensed gases  the ground state kinetic energy is strongly quenched
by interactions and, for large $N$, the ratio $E_{int}/E_{ho}$ is
equal to $2/3$ (see eq.(\ref{virial})). In this limit one recovers the results 
$\omega_M=\sqrt5\omega_{ho}$ and $\omega_Q=\sqrt2\omega_{ho}$
predicted by the hydrodynamic theory of superfluids
\cite{ss}. Conversely, in the Fermi case 
the interaction energy is, in most cases, only
a small perturbation that can be safely estimated using the
Thomas-Fermi expression \cite{silvera_et_al}
\begin{equation} \label{density}
\rho(r)=\left(\frac{2m}{\hbar^2} \right)^{3/2} \frac 1
{3\pi^2} \left(\mu_0-V_{ext}(r) \right)^{3/2}
\end{equation}
for the ground state density. In this equation
$\mu_0=(3N)^{1/3}\hbar \omega_{ho}$ is  the
chemical potential of the Fermi gas, fixed by the normalization condition.
Using expression (\ref{density}) for the density one obtains the result
\begin{equation} \label{ratio}
{E_{int}\over E_{ho}} = \alpha \frac{N^{1/6}a}{a_{ho}}
\end{equation}
where $\alpha = 8192\ \sqrt 2\ 3^{1/6}/2835\pi^2 \simeq 0.50$ while $a_{ho}=
(\hbar/m\omega_{ho})^{1/2}$ is the usual  harmonic oscillator
lenght. Interaction effects  are governed by the 
combination $N^{1/6}a/a_{ho}$ showing that in order to emphasize the
role of interactions it is much more efficient to increase the ratio $a/a_{ho}$
rather than the number of atoms. The ratio (\ref{ratio}) can be also casted
in the form $E_{int}/E_{ho}\simeq 0.3\times k_Fa$ where $k_F=(2m (3N)^{1/3}
\omega_{ho}/\hbar)^{1/2}$ is the Fermi momentum of the trapped gas
\cite{silvera_et_al}. 

In analogous way one can calculate  the frequency of the
out of  phase oscillations, hereafter called spin excitations. We
report here the result for the most relevant spin dipole mode
excited by the operator $F=\sum_{i \uparrow} z_i-\sum_{i
\downarrow}z_i$. 
This mode corresponds to a relative oscillation of the centers of mass
of the spin-up 
and spin-down clouds and is the analog of the giant dipole resonance  
exhibited by atomic nuclei \cite{bm}. The frequency of the spin dipole
mode is found to be
\begin{multline} \label{SD}
\omega_{SD} =\sqrt{\omega_{ho}^2-\frac g {mN}\int |\partial_z
  \rho|^2\,d\er}  
\\ \simeq \omega_{ho} \left(1-\alpha' \frac{N^{1/6}a}{a_{ho}}\right)
\end{multline}
where $\alpha'=128\ \sqrt 2\ 3^{1/6}/35 \pi^2 \simeq 0.63$ and,
 in the second equality, we have used 
the ground state density
(\ref{density}) to evaluate the interaction contribution to first order
in the scattering length.
Notice that interactions affect the spin-dipole mode through the same
combination of  parameters characterizing  the ratio (\ref{ratio}). 
Typical values for the fermionic isotope of potassium 
show that interaction effects are rather small. For example, using
$N=5\times 10^5$ and
$a/a_{ho} = 6\times 10^{-3} $ one finds that the dipole frequency decreases by
$\sim 3 \%$.
Significantly larger corrections are  predicted in the case of $^6$Li where
the scattering length is a factor 10 larger and negative. Notice that
the integrals  
characterizing the interaction contribution to the collective frequencies,
are calculated here at $T=0$ and  would decrease  at higher
temperature. 
We also note that, at $T=0$, sum rules provide only an upper
bound to the frequency of the lowest state excited by the operator
$F$. However, if one considers first
order corrections in the interaction constant $a$,  the sum rule bounds
(\ref{M}-\ref{D},\ref{SD}) can be shown to correspond to 
the exact value of the collective frequencies.

The results derived above can be easily generalized to the case of
deformed traps in which the external potential reads
$V_{ext}({\bf{r}})=m\omega_{\perp}^2 (x^2+y^2+\lambda^2z^2)/2$ and
$\lambda=\omega_z/\omega_{\perp}$ is the deformation parameter of the trap.
By assuming that interaction effects are smaller than the unperturbed
splitting between the radial and axial frequencies, one  finds simple
results also in  this case.
The new decoupled frequencies, associated with the radial and axial 
excitation operators 
$F_{radial}=\sum_i x_i^2+y_i^2$ and $F_{axial}=\sum_i z_i^2$, become,
to first order in $a$,
$\omega_{radial} = 2\omega_{\perp}$
and 
$\omega_{axial} = 2 \omega_z (1- 3 /16\ \alpha N^{1/6}a/a_{ho})$
showing that only the axial mode is affected by the interaction.
The oscillator length $a_{ho}=(\hbar/ m \overline{\omega})^{1/2}$ is here
defined in terms of the geometrical average $ \overline{\omega}=
\omega_\perp \lambda^{1/3}$ of the three frequencies. 
For the spin dipole mode result (\ref{SD}) is easily 
generalized to both the radial and axial directions. One finds
$\omega_{radial} = \omega_{\perp} (1-\alpha' N^{1/6}a/a_{ho})$
and $\omega_{axial} = \omega_z (1-\alpha' N^{1/6}a/a_{ho})$.

So far we have ignored the effects of collisions. If the collisional 
frequency
is much larger than the frequency of the collective excitations then the system
is in the hydrodynamic regime, also known as first sound regime. 
An important question is 
whether the transition between the collisionless and hydrodynamic
regimes takes place in the degenerate or classical regime for our
trapped gases. Let us first discuss the collective
frequencies of a trapped Fermi gas in the full hydrodynamic regime
(see \cite{tosi} for a recent discussion). 
In the spherical case 
the quadrupole frequency becomes $\omega_Q^{HD} = \sqrt 2 \omega_{ho}$
instead of (\ref{Q}). This result is independent of statistics
and holds also for a classical gas \cite{griffin}. The reduction with respect
to the  collisionless value  is due to the fact that,  in the
collisional regime, the only restoring force for surface excitations
arises from the external field. 
For the monopole frequency one instead finds that result (\ref{M}) 
holds also in the hydrodynamic regime. The above analysis
can be also extended to the case of deformed traps \cite{tosi} where one finds
that, in the absence of mean
field effects, the collective frequencies coincide 
with the ones holding for a classical gas \cite{griffin}.

Collisions are expected to have more dramatic consequences
on the spin-dipole oscillation since they do not
conserve the spin current and
consequently give rise, in the hydrodynamic regime, to a pure diffusive mode. 
The investigation  of
the spin dipole mode is consequently expected to be  a 
sensitive test of
the role of collisions and, possibly, of quantum statistics.
The equations for the spin dipole oscillation can be easily obtained 
starting from the Boltzmann equation where the effects of Fermi statistics
are included in the collisional integral. For a first estimate let us ignore
the mean field effect, which is responsible for the frequency shift
of (\ref{SD}). Using the method of the averages recently developed in
\cite{david} for classical trapped gases, one obtains the
following coupled equations: 
\begin{align}
\label{SD+collisions1}
&\partial_t \langle z_{\uparrow}-z_{\downarrow} \rangle
- \langle v_{z \uparrow}-v_{z \downarrow} \rangle=0\\ 
\label{SD+collisions2}
&\partial_t \langle v_{z \uparrow}-v_{z\downarrow}
  \rangle+ \omega^2_z
\langle z_{\uparrow}-z_{\downarrow}  \rangle = \langle (v_{z
  \uparrow}-v_{z \downarrow} ) I_{coll} \rangle
\end{align}
where the average $\langle \ldots\rangle$ is taken both in coordinate and
momentum space.
The collisional term is given by
\begin{multline}
\label{collision}
\langle (v_{z\uparrow}-v_{z\downarrow})  I_{coll}
\rangle=\\-\frac{\sigma m^6} {4\pi h^6}\frac 2 N \int d\er 
d\vu _1 d\vu _2 d \Omega |\vu_1-\vu_2| (v_{1 z}-v_{2 z}) \times \\ \times
[ (1-f_{1 \uparrow})(1-f_{2 \downarrow})f_{1
    \uparrow}' f_{2\downarrow}' - f_{1\uparrow}f_{2
    \downarrow}(1-f_{1\uparrow}')(1-f_{2\downarrow}')]
\end{multline}
where $\sigma=4\pi a^2$ is the total cross section, $\vu'_1$ and $\vu'_2$
are the velocities of the particles $1$ and $2$ after the collision,
and $f_\uparrow$, $f_\downarrow$ are the distribution functions
relative to the two spin components, normalized to $(m/h)^3 \int
f_\uparrow  d\er d\vu=(m/h)^3 \int
f_\downarrow d\er d\vu= N/2$.
The collisional term can be 
estimated by assuming that during the oscillation the
distribution functions of the two spin species behaves, 
in velocity space, as
$f_{\uparrow \downarrow}(v_x,v_y,v_z)= f_0(v_x,v_y,v_z \pm u)$, 
where $f_0$ is the distribution function of each component 
at thermal equilibrium. This corresponds to a rigid displacement of
the velocity Fermi distributions of the two spin components in
opposite directions. One finds 
$ \langle v_{z \uparrow}-v_{z \downarrow} \rangle =2u $ and,
by developing the integral (\ref{collision}) 
to first order in $u$, one can finally write
\begin{equation}
\label{relaxationtime}
\langle (v_{z\uparrow}-v_{z\downarrow})  I_{coll}
\rangle= - {\langle v_{z\uparrow}-v_{z\downarrow}\rangle \over \tau}
\end{equation}
where
\begin{multline} 
\label{tau}
\frac 1 {\tau} =  \frac{\sigma m^7}{24\pi h^6NK_BT}\int d\er 
d\vu _1 d\vu_2 d \Omega |\vu| (\vu -\vu' )^2 \times \\ \times
f_0(\vu_1)f_0(\vu_2)(1-f_0(\vu'_1))(1-f_0(\vu'_2)) 
\end{multline}
defines the relevant relaxation time of the spin dipole
oscillation. In (\ref{tau}) we have defined $\vu=\vu_1-\vu_2$ and
$\vu'=\vu'_1-\vu'_2$. 
As a consequence of (\ref{relaxationtime}) the equations of motion
(\ref{SD+collisions1}-\ref{SD+collisions2}) for the
spin dipole oscillation take the simple form of a
damped harmonic oscillator. Looking for solutions of the form $e^{-i
  \omega t}$ the dispersion law is given by
\begin{equation} \label{damping}
 \omega=-\frac 1 {2 \tau} \left(i \pm  \sqrt{4 \omega_z^2 \tau^2-1} \right)
\end{equation}
showing that the oscillations become overdamped  if $\omega_z\tau < 1/2$.

\begin{figure}
\begin{center}
\includegraphics[width=0.4\textwidth]{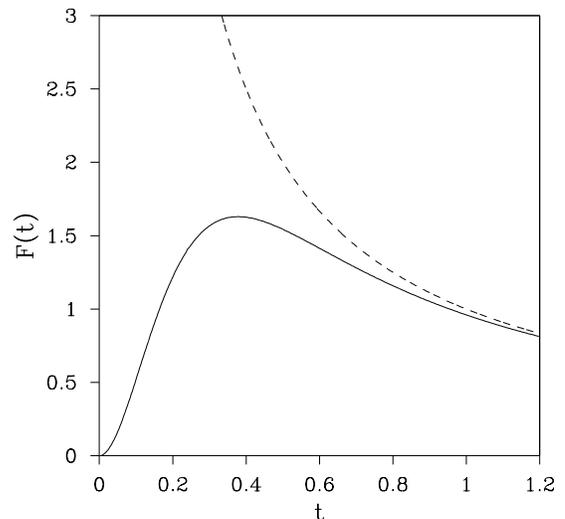}
\caption{Spin dipole relaxation function (see eq.(\ref{mainresult}))
  as a function of the reduced temperature $t=T/T_F$. The classical
  prediction $1/t$ (dahsed line) is also shown.}
\label{fig1}
\end{center}
\end{figure}

The relaxation time (\ref{tau}) is easily evaluated at high temperature
where the effects of Fermi statistics are negligible. In this case
it takes the form
\begin{equation}
{1\over \tau_{cl}} = \frac 2 3 \gamma_{cl}
\label{gamma}
\end{equation}
where $\gamma_{cl} = v_{th} \sigma \rho(0)/2$ is the classical collisional
rate, $v_{th}=(8 K_BT/\pi m)^{1/2}$ is the thermal velocity and
$\rho(0)=\rho_{\uparrow}(0)+ \rho_{\downarrow}(0)$ is the central
density of the gas.  At  
temperatures smaller than the Fermi temperature $T_F=(3N)^{1/3}\hbar
\omega_{ho}/K_B$  
the effects of statistics become important and, for $T \to 0$, the
relaxation time becomes larger and larger exhibiting the typical behaviour 
of Fermi systems. A similar behaviour has been recently shown to
occur in the relaxation of the motion of a
classical particle inside a degenerate trapped Fermi gas \cite{ferrari}.
By a proper change of the variables entering the collisional integral
(\ref{tau}), the dimensionless quantity $\omega_z\tau$ can be written
in the useful form (for oscillations along the radial direction one should
consider the quantity $\omega_\perp \tau=\omega_z\tau/\lambda$)
\begin{equation}
{1\over \omega_z\tau} = \frac 4
{3^{4/3} \pi} \lambda^{-2/3}\left(N^{1/3}{a\over a_{ho}}\right)^2
F\left(\frac T {T_F} \right)
\label{mainresult}
\end{equation}
where $F(t)$ is a dimensionless function, sensitive to Fermi statistics,
which determines the temperature dependence of the relaxation time.
This function is plotted in Fig.(\ref{fig1}). For large values of the reduced 
temperature $t=T/T_F$ it approaches the classical behaviour $1/t$,
while for $t$ smaller than $1$ one observes, as expected, important
deviations due to quantum effects. At very low temperatures the
function $F(t)$ can also be calculated analytically and we find
\begin{equation} \label{lowt}
F(t) = 8 \pi^2 t^2.
\end{equation}
Of course result (\ref{lowt}) holds above the superfluid transition.
The figure shows that the system
will be  in the collisionless regime both 
at sufficiently high and low temperatures.
In the first case the gas is classical and collisions are rare because
the density is very low. In the second one the gas is degenerate and collisions
are rare because of Fermi statistics. The possibility for the system
to reach the overdamped regime $\omega_z\tau < 1/2$
depends in a crucial way on the value of the physical parameters
entering (\ref{mainresult}). The maximum value  of $F(t)$, 
reached at $t\sim 0.4 $, is $\sim 1.6$ 
so that the condition for the existence of overdamping is given by 
$\lambda^{1/3}N^{1/3}| a |/a_{ho} > 2.1$
(for the spin dipole oscillation along the radial axis the condition is
instead $\lambda^{-1/6}N^{1/3}| a |/a_{ho} > 2.1$).
This condition should be easily
achievable in the case of lithium. For example, choosing a spherical trap 
with $N=5\times 10^5$ and 
$ a /a_{ho} = -3\times 10^{-2}$ one finds overdamping for 
$0.2 < T/T_F <0.8 $.
For lower temperature
the system will exhibit a transition to the collisionless
regime where the spin oscillation is damped according to 
eq.(\ref{damping}). In the case of $^{40}$K, where
the value of the scattering length is much smaller, the system is
expected to be
always far from the hydrodynamic regime.
Nevertheless an accurate investigation of the damping  
might point out the effects of Fermi statistics also in this case. In
fact the damping of the spin oscillation is predicted to strongly decrease at
low temperature in contrast to the classical behaviour.
Notice that the above estimates are based on the unperturbed value $\omega_z$
($\omega_\perp$) for the spin dipole frequency. At low  temperature one
can easily renormalize these values by including the mean field effect 
through  eq.(\ref{SD}).

In conclusion, we have provided a systematic discussion of
interactions effects on the propagation of collective oscillations in
trapped Fermi gases. In the collisionless regime interactions
renormalize the collective frequencies through mean field
effects. The predicted shifts should be visible in experiments due to the
high precision of frequency measurements. Collisions are responsible
for the damping of the oscillations and their effect turns out to be
very sensitive to Fermi statistics. Overdamped oscillations are
predicted to occur in the case of the spin dipole oscillation and
should be visible in lithium due to
the high value of the scattering lenght. The
present analysis can be naturally extended to include asymmetric
configurations with different numbers of atoms in the
two spin states ($N_\uparrow \neq N_\downarrow$) and/or different
trapping potentials ($V^\uparrow_{ext} \neq V^\downarrow_{ext}$).

Useful discussions with D.\ Gu\'ery-Odelin, H.\ Stoof and F.\ Zambelli
are acknowledged.

\begin{thebibliography}{11}
\bibitem{bec} M. H. Anderson \emph{et al.}
, Science {\bf 269} 198 (1995);
  K. B. Davis \emph{et al.}
, Phys. Rev. Lett. {\bf75}, 3969  (1995); 
C.\ C.\ Bradley \emph{et al.}
 , Phys.\ Rev.\ Lett {\bf 78}, 985 (1997); see also Phys.\ Rev.\ Lett.\
  \textbf{75}, 3969 (1995)
\bibitem{rmp} F.\ Dalfovo, S.\ Giorgini, L.\ P.\ Pitaevskii and S.\
  Stringari, Rev.\ of Mod.\ Phys.\ {\bf 71} 463 (1999)
\bibitem{exp} D.\ S.\ Jin \emph{et al.}, Phys.\ Rev.\ Lett.\
  \textbf{77}, 420 (1996); M.-O.\ Mewes \emph{et al.}, Phys.\ Rev.\
  Lett.\ \textbf{77}, 988 (1996); D.\ M.\ Stamper-Kurn, H.-J.\
  Miesner, S.\ Inouye, M.\ R.\ Andrews, W.\ Ketterle, Phys.\ Rev.\
  Lett. \textbf{81}, 500 (1998)
\bibitem{houbiers} M.\ Houbiers, R.\ Ferwerda, H.\ T.\ C.\ Stoof, W.\ I.\
  McAlexander, C.\ A.\ Sackett, R.\ G.\ Hulet, Phys.\ Rev.\ A
  \textbf{56}, 4864 (1997)
\bibitem{theo-fermi} G.\ M.\ Bruun, K.\ Burnett, Phys.\ Rev.\ A
  \textbf{58}, 2427 (1998)
\bibitem{molmer} K.\ M{\o}lmer, Phys.\ Rev.\ Lett.\ \textbf{80}, 3419 (1998)
\bibitem{tnpi1} M.\ Amoruso, A.\ Minguzzi, S.\ Stringari, M.\ P.\
  Tosi, L.\ Vichi, Eur.\ Phys.\ J. D \textbf{4}, 261 (1998)
\bibitem{exp-fermi} M.\ Prevedelli \emph{et al.}
, Phys.\ Rev.\ A \textbf{59}, 886 (1999); B.\ DeMarco, J.\
  L.\ Bohn, J.\ P.\ Burke Jr.\, M.\ Holland, D.\ S.\ Jin,
  \emph{cond-mat}/9812350 
\bibitem{kagan} M.\ A.\ Baranov, Yu.\ Kagan, M.\ Yu.\ Kagan, JETP
  Lett.\ \textbf{64}, 301 (1996)
\bibitem{stoof} H.\ T.\ C.\ Stoof, M.\ Houbiers, C.\ A.\ Sackett, R.\
  G.\ Hulet, Phys.\ Rev.\ Lett.\ \textbf{76} 10 (1996)
\bibitem{petrov} M.\ A.\ Baranov, D.\ S.\ Petrov,
  \emph{cond-mat}/9901108  
\bibitem{pines} D.\ Pines, P.\ Nozi\'eres \emph{The Theory of
  Quantum Liquids} Vol.\ I, Addison-Wesley
\bibitem{lipparini} S.\ Stringari, J.\ Low Temp.\ Phys.\ \textbf{57},
  307 (1984); O.\ Bohigas, A.\ M.\ Lane, J.\ Martorell,
  Phys.\ Rep.\ \textbf{51}, 267 (1979); E.\ Lipparini and S.\
  Stringari, Phys.\ Rep.\ {\bf 175} (1989) 103-261
\bibitem{ss} S.\ Stringari, Phys.\ Rev.\ Lett.\ \textbf{77}, 2360 (1996)
\bibitem{silvera_et_al} I.\ F.\ Silvera, Physica \textbf{109 \& 110B}
  1499 (1982); D.\ A.\ Butts, D.\ S.\ Rokhsar, Phys. Rev. A {\bf 55}
4346 (1997)
\bibitem{bm} A.\ Bohr, B.\ R.\ Mottelson, \emph{Nuclear structure}
  Vol.\ II, Benjamin
\bibitem{tosi} M.\ Amoruso, I.\ Meccoli, A.\ Minguzzi, M.\ P.\ Tosi,
  accepted for pubblication in Eur. Phys. J. D 
\bibitem{griffin} A.\ Griffin, Wen-Chin Wu, S.\ Stringari, Phys.\
  Rev.\ Lett.\ {\bf78} 1838 (1997)
\bibitem{ferrari} G.\ Ferrari, \emph{cond-mat}/990416, accepted for
  pubblication in Phys.\ Rev.\ A, rapid communications
\bibitem{david} D.\ Gu\'ery-Odelin, F.\ Zambelli, J.\ Dalibard, S.\
  Stringari, preprint \emph{cond-mat}/9904409
\end{thebibliography}
\end{document}